# 3-D PET Image Generation with tumour masks using TGAN

Robert V Bergen*[a], Jean-Francois Rajotte[a], Fereshteh Yousefirizi[b], Ivan S Klyuzhin[b,c], Arman Rahmim[b,c], Raymond T. Ng[a]

[a]University of British Columbia Data Science Institute, 6339 Stores Road, Vancouver, BC, Canada V6T 1Z4; [b]Department of Integrative Oncology, BC Cancer Research Institute, Vancouver, BC, Canada; [c]Department of Radiology, University of British Columbia, Vancouver, BC, Canada

**ABSTRACT**

Training computer-vision related algorithms on medical images for disease diagnosis or image segmentation is difficult due to the lack of training data, labeled samples, and privacy concerns. For this reason, a robust generative method to create synthetic data is highly sought after. However, most three-dimensional image generators require additional image input or are extremely memory intensive. To address these issues we propose adapting video generation techniques for 3-D image generation. Using the temporal GAN (TGAN) architecture, we show we are able to generate realistic head and neck PET images. We also show that by conditioning the generator on tumour masks, we are able to control the geometry and location of the tumour in the generated images. To test the utility of the synthetic images, we train a segmentation model using the synthetic images. Synthetic images conditioned on real tumour masks are automatically segmented, and the corresponding real images are also segmented. We evaluate the segmentations using the Dice score and find the segmentation algorithm performs similarly on both datasets (0.65 synthetic data, 0.70 real data). Various radionomic features are then calculated over the segmented tumour volumes for each data set. A comparison of the real and synthetic feature distributions show that seven of eight feature distributions had statistically insignificant differences ($p > 0.05$). Correlation coefficients were also calculated between all radionomic features and it is shown that all of the strong statistical correlations in the real data set are preserved in the synthetic data set.

## 1. INTRODUCTION

Developing a deep-learning based disease diagnosis system using medical images is a very data intensive task due to the large number of parameters that must be learned. In a supervised learning setting, annotated samples are also required which demand expert knowledge of the specific dataset. Data sharing between institutions is one way to overcome data requirements, but can be difficult due to privacy concerns[1]. As such there are very few public medical datasets available online, and those that are available may vary in quality. The studies may also be imbalanced since healthy subjects are easier to find and recruit.

One possible solution to the data problem is federated learning, whose use in healthcare has been growing in recent months; Sheller et al. are among the first researchers who applied federated learning to solve a medical imaging problem, namely to segment brain tumors[2]. However, federated learning has several limitations: 1) It has been shown that removing the patient metadata alone does not guarantee data privacy, e.g. the original training data from individual institutions can be reconstructed by probing the outputs and structure of the network that was trained in a federated setting. 2) Certain jurisdictions may not allow the use of medical images for cross-institutional model training, even with complete anonymization, patient consent and institutional approval. 3) A separate legal agreement may be required for each entity interested in using imaging data from a particular institution. 4) Data from different institutions may be highly heterogeneous and/or not independent and identically distributed[3,4]. An alternative solution to the data sharing bottleneck that we explore here is based on synthetic image generation. Since many studies often analyze data on the level of cohorts rather than individuals, synthetic images with similar properties (feature distributions and co-distributions) to real images can be used to share valuable information about the original dataset, without compromising any sensitive patient information.

*robert.bergen@ubc.ca; phone 1 (604) 822-5852; https://dsi.ubc.ca

Generative Adversarial Networks (GANs) represent a technique capable of generating high-quality medical images[5,6]. GANs consist of a generator and a discriminator which play an adversarial game; the GAN tries to generate images to trick the discriminator, while the discriminator tries to distinguish between the fake and real samples. GANs have shown promise in several medical imaging studies; for example, they have been used to generate synthetic abnormal MRI images with brain tumors, to synthesize high-resolution retinal fundus images and generate synthetic pelvic CT images[7,8,9]. Specifically within the PET imaging domain, GANs have also been used to generate synthetic 2-D brain images[10].

Rather than generating standalone PET images, the majority of GAN-based PET image generators in the literature focus on image-to-image translation applications[11]. That is, the generator will be given one type of image as input, and output a translated image. For example, GANs may be trained to convert low-dose PET to full-dose PET, or to convert PET images to CT images[12,13]. However, generating standalone PET images is more complicated because it is very time consuming for the generator to learn the spatial properties of a 3-D medical image data set without having any additional image inputs as guidance, and requires a large training dataset.

To address the issue of computational efficiency, we turn to video generation techniques as inspiration. Video GANs have the same fundamental problem in that their data is three-dimensional and would require an enormous amount of memory to train using 3-D convolutional layers. Many video generation techniques approach the 3-D problem by attempting to predict a future video frame based on the current frame. Some proposed architectures decompose the frames into foreground and background, which is a method that may be ill-suited for 3-D medical image generation[14]. For this reason, we choose to adopt Temporal GANs (TGANs), which use a generator consisting of two sub-networks: a temporal generator and an image generator[15]. The image generator generates 2-D frames while the temporal generator captures the dynamics of the system and how the frames evolve over time. While the discriminator of TGAN is fully 3-D, the computational demand of 3D video generation is partially addressed by substituting a 3-D generator for the 1-D and 2-D generators. TGAN can easily be adapted for 3-D PET image generation, by substituting the frames of a video for a series of 2-D image slices of the 3-D PET volume.

In our experiment, we show that we are able to generate high-quality 3-D head and neck PET images using the TGAN architecture. We also modify the TGAN architecture to accept additional conditional input in the form of a binary tumour mask. The tumour mask can be used to specify the size, shape and location of a generated tumour, which can be used as a ground truth for downstream image segmentation tasks.

In this work, we chose to use image segmentation as a measure of fidelity and utility of the synthetic data set. We train a segmentation model on real data and test its performance on synthetic data. The rationale is that the segmentation algorithm should perform similarly on high quality synthetic data compared to real data. We are also able to use the segmentation to perform further analysis on the real and synthetic segmented volumes by calculating various radionomic features. We compare the radionomic feature distributions of real and synthetic data sets and evaluate whether strong statistical correlations in the real data set are preserved in the synthetic data set. Finally, we perform a data augmentation experiment to evaluate the utility of our synthetic data in cases where only a small real dataset is available.

## 2. METHODOLOGY

### 2.1 Unconditional TGAN

The unconditional TGAN consists of a temporal generator, $G_0$, and an image generator, $G_1$. The temporal generator takes a random latent variable $z_0$ as input and generates a temporal latent vector $z_1(t)$. The image generator generates frames of a video at time $t$ using the $z_0, z_1(t)$ variables as input. That is, a $T$-frame video is represented as the time series $[G_1(z_0, z_1(t=1)), ..., G_1(z_0, z_1(t=T))]$. To improve stability of the training process, the spectral norm of each weight parameter in each layer is constrained to less than 1, which Saito et al. refer to as singular value clipping[15].

In our experiment we use 3D head and neck PET images, substituting the time dimension in videos for the 3$^{rd}$ spatial (axial) dimension in the 3D volumes.

## 2.2 Conditional TGAN

We make a modification to the original TGAN architecture by conditioning the image generator/discriminator pair on the ground truth tumour masks, allowing us to generate a 3-D volume with specific tumour geometry. In the temporal generator, we pass the tumour mask into additional convolutional layers and then concatenate this vector with the other latent vectors $z_0, z_1$, before deconvolution. In the discriminator, we simply add the tumour mask as another input image channel before convolution. We also introduce a new hyperparameter, $\omega$, to the model, which determines how strongly the conditional tumour mask information is weighted relative to the input image in the discriminator. The weighting of image $I$ and mask $M$ is given by $I' = (1 - \omega)I$, $M' = \omega M$. Both the unconditional and conditional TGAN models were trained for 5000 epochs using the RMSProp optimizer with learning rate 0.00005, and Wasserstein loss on a P40 GPU (24 GB) with a batch size of 32. We perform singular value clipping every 5 iterations, which is the recommended frequency in the original TGAN paper[15].

## 2.3 Segmentations

To segment our real and synthetic tumours, we use the same neural network segmentation architecture as the MICCAI 2020 Head and Neck Tumor (HECKTOR) segmentation challenge winner, described by Iantsen et al[16]. This network is designed on the U-net architecture with residual layers and supplemented with squeeze-and-excitation normalization. The only modification we make is the number of inputs. The original architecture takes in two inputs; the head & neck PET image as well as a CT image. Our version only accepts PET images as input. The models were trained for 150 epochs using a P40 GPU (24 GB) with a batch size of 2. Segmentation quality was evaluated using the DICE score (DSC) metric.

## 2.4 Image features

We choose a number of common radionomic image features to calculate over the real and synthetic data sets. They are the metabolic tumour volume (MTV), the mean, max, and peak of the standardized uptake value ($SUV_{mean}$, $SUV_{max}$, $SUV_{peak}$), total lesion glycolysis (TLG), and gray level co-occurrence matrix (GLCM) features (GLCM Entropy, GLCM Energy, and GLCM Homogeneity). This mix of features represents both first order measurements of voxel intensities and second-order measurements that represent heterogeneity and texture[17]. We evaluate the similarity between real image features and synthetic data features sets by performing t-tests on their distributions with a significance level $\alpha = 0.05$. We also calculate the correlation coefficients between image features to ensure that strong correlations of radionomic features in the real data set are preserved in the synthetic data set.

## 2.5 Data Augmentation

A data augmentation experiment was performed by training a segmentation model, S, on real data from one out of four centers in the HECKTOR dataset (CHGJ). A TGAN was then trained on the remaining HECKTOR dataset and synthetic data was generated to augment S. The performance of S was evaluated on two validation sets; one comprises a small subset of the CHGJ center's data and another validation set comprises a random selection of samples from outside the CHGJ center. Segmentation performance on these sets indicate the utility of the synthetic images on segmentations performed within the CHGJ center and images outside CHGJ, respectively.

## 3. DATA

We utilized a publicly available dataset in The Cancer Imaging Archive (TCIA), further refined within the MICCAI 2020 Head & Neck Tumor (HECKTOR) challenge[18]; it comprises 201 cases from four centers. Each case comprises a PET image and GTVt (primary Gross Tumor Volume) mask, as well as a bounding box location. We use the bounding box information to crop the PET and GTVt masks to 64x64x32 volumes for input into the TGAN and segmentation networks. The in-plane (axial) resolution of the PET images ranged from 3.5 mm to 3.9 mm while the through-plane resolution was 3.7mm. After cropping, this corresponds to a minimum field of view of (224 mm × 224 mm × 118.4 mm). For unconditional TGAN, all 201 cases were used for training. For conditional TGAN, 11 cases were randomly withheld for testing. For training the segmentation neural networks, 25% of the cases were randomly withheld for testing.

# 4. RESULTS

## 4.1 Synthetic Data

Representative samples of our synthetic images generated by the unconditional TGAN are shown in Figure 1. We show samples of synthetic images generated by the conditional TGAN in Figure 2. Each synthetic image is shown with the tumour mask overlaid in red. We also show the corresponding real PET image with tumour mask overlaid for comparison.

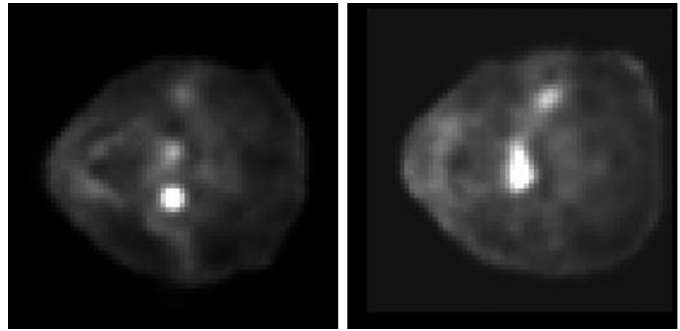

Figure 1. Two representative examples of synthetic PET images generated by the unconditional TGAN. Both images are shown with the same window and level settings.

The PET images generated by both the unconditional and conditional TGANs are volumes of size 64x64x32. The unconditional TGAN is able to generate images that appear to have lesions (Figure 1). Since we have no ground truth for tumours in the unconditional images, our analysis at this stage is only qualitative and based on visual inspection. We observed that the shape of the head and internal anatomy structure all appeared realistic. The conditional TGAN faithfully reproduces lesions based on the input tumour mask (Figures 2,3). Figure 2 shows images generated by masks that were seen by the generator during training, while Figure 3 shows images generated by masks that were not seen during training. It appears that there is good generalization and there are no visibly noticeable issues when generating images given unseen tumour masks.

The hyperparamater $\omega$ was introduced into the conditional TGAN to produce more realistic anatomy. Figure 4 shows an example of two synthetic images generated with the same tumour mask, but with different $\omega$ values. At $\omega = 1$, the hyperparameter has no effect on the model. Upon visual inspection, the synthetic anatomy at $\omega = 1$ can be very unrealistic, producing unnaturally shaped heads. Reducing $\omega$ to 0.01 forces the generator to weigh input images more strongly than input masks during training, thereby generating more realistic anatomy overall. Although the same tumour mask was used, the lesions generated are slightly different in the two examples. This is due to the fact that one model weights the input tumour mask much more strongly than the other.

Interestingly, TGAN was originally developed for 16-frame videos, beyond which the training would become unstable[12]. We are able to achieve stable training at a larger number of frames (32), likely because the general structure of PET images are more predictable than the typical structure of video data sets.

Theoretically, TGAN could be used to generate higher-resolution volumes and our initial tests show that training is also stable at 64x64x64 volumes. We chose to crop our images at 64x64x32 because of memory constraints for our system and long training times for larger volumes. Despite the small volume size, we are still able to encapsulate the entire head and neck for each patient with this crop setting.

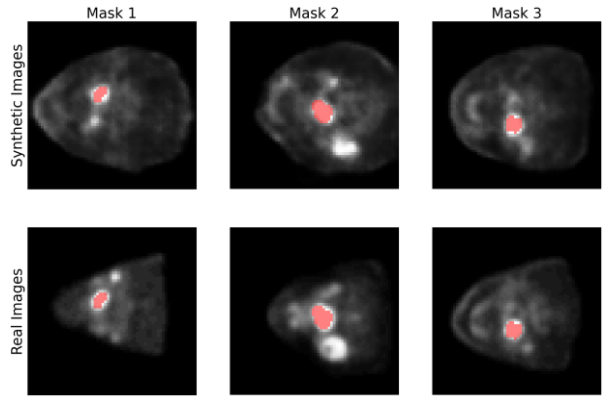

Figure 2. Examples of real and synthetic PET images with tumour masks overlaid in red. Each column shows the synthetic image generated by a tumour mask and the corresponding real PET image. The tumour masks shown here were included in the training set.

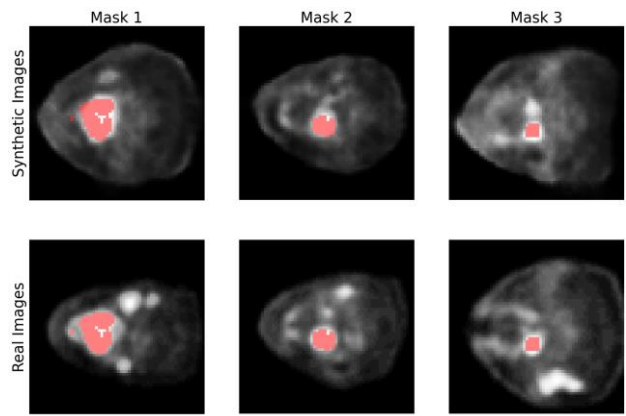

Figure 3. Examples of real and synthetic PET images with tumour masks overlaid in red. Each column shows the synthetic image generated by a tumour mask and the corresponding real PET image. The tumour masks shown here were not included during training.

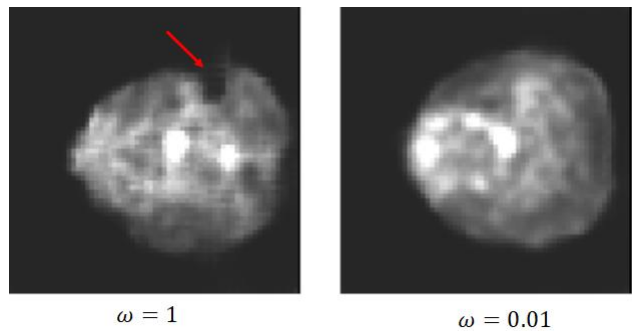

Figure 4. Synthetic PET images generated by the conditional TGAN with the same tumour mask, using different $\omega$ values. On the left, the contour of the head is not smooth and a large cavity is visible (red arrow). Reducing $\omega\ to$ 0.01 produces a much smoother contour while maintaining a realistic lesion. Add mean/max intensities inside and outside mask

### 4.2 Fidelity: Segmentations

We generated 200 synthetic volumes on both the unconditional and conditional TGAN, then performed automatic segmentations on them. Since there is no ground truth available for the unconditioned synthetic images, a Dice score cannot be calculated. However, the Dice scores for conditioned synthetic images were calculated and compared to real values.

The average Dice score for real and synthetic data was 0.7 and 0.65, respectively. In Figure 5, the distribution of Dice scores are shown for both data sets. As a reference point, Iantsen et al achieved a 0.759 Dice score when using additional CT data for guidance[16]. We observed only a small drop in segmentation performance on our real data when supplementary CT images are not used. Synthetic Dice scores distributions were similar to real Dice score distributions. One immediately observable difference between the Dice score distributions is the number of 0.0 Dice score segmentations. We observed that these particular synthetic images were trained on tumour masks that were small in volume. The model faithfully reproduces lesions based on the input tumour masks, but may produce other lesions elsewhere in the image which the segmentation algorithm misidentifies as the primary lesion. Figures 6 shows some examples of real and synthetic tumour segmentations, along with the original masks used to synthesize the images. Real and synthetic data are displayed at identical window and level settings.

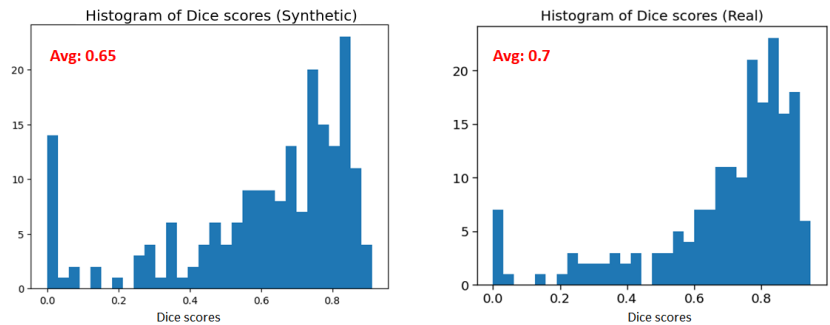

Figure 5. Histograms of Dice scores for the synthetic and real datasets ($n = 200$). The average dice scores are shown in red.

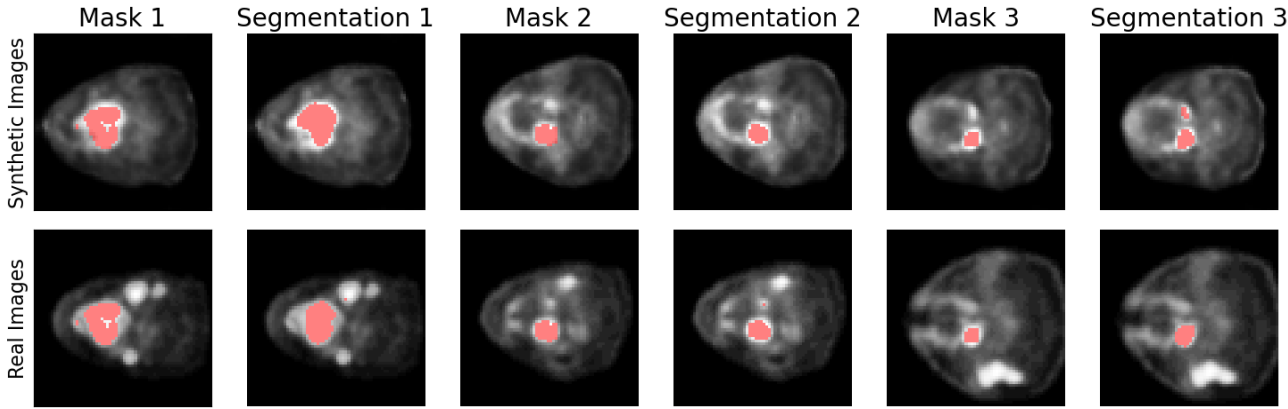

Figure 6. Fidelity evaluation. Columns 1, 3, 5: Randomly selected examples of real and synthetic PET images with tumour masks overlaid in red. These columns show the synthetic image generated by a tumour mask and the corresponding real PET image. Columns 2, 4, 6: Examples of tumour segmentations for real and synthetic PET images.

### 4.3 Fidelity: Image features

Pair plots of radiomic features calculated over the real and synthetic tumour segmentations are shown in Figures 7 and 8. Figure 7 shows the pair plots for the synthetic images generated by the unconditional TGAN and Figure 8 shows the pair plot for the synthetic images generated by the conditional TGAN. T-tests were performed on the feature distributions and show that for the unconditional TGAN, the GLCM Entropy, Energy, and Homogeneity features were significantly different from each other (p <0.05). Using conditional TGAN, only the GLCM energy distributions were significantly different from each other (p <0.05).

To show that the conditional TGAN synthetic images preserve strong correlations between the features, we calculated the correlation coefficient for each feature pair in the real data set and list the largest magnitude correlation coefficient in Table 1. The same correlation coefficients on the synthetic data set are presented for comparison. For most correlation coefficients, we observed less than a 25% percent difference between real and synthetic data. The only exception was a weak negative correlation between GLCM Energy and GLCM Entropy we observed in the real data (r = -0.4), which was not observed in the synthetic data (r = 0.03).

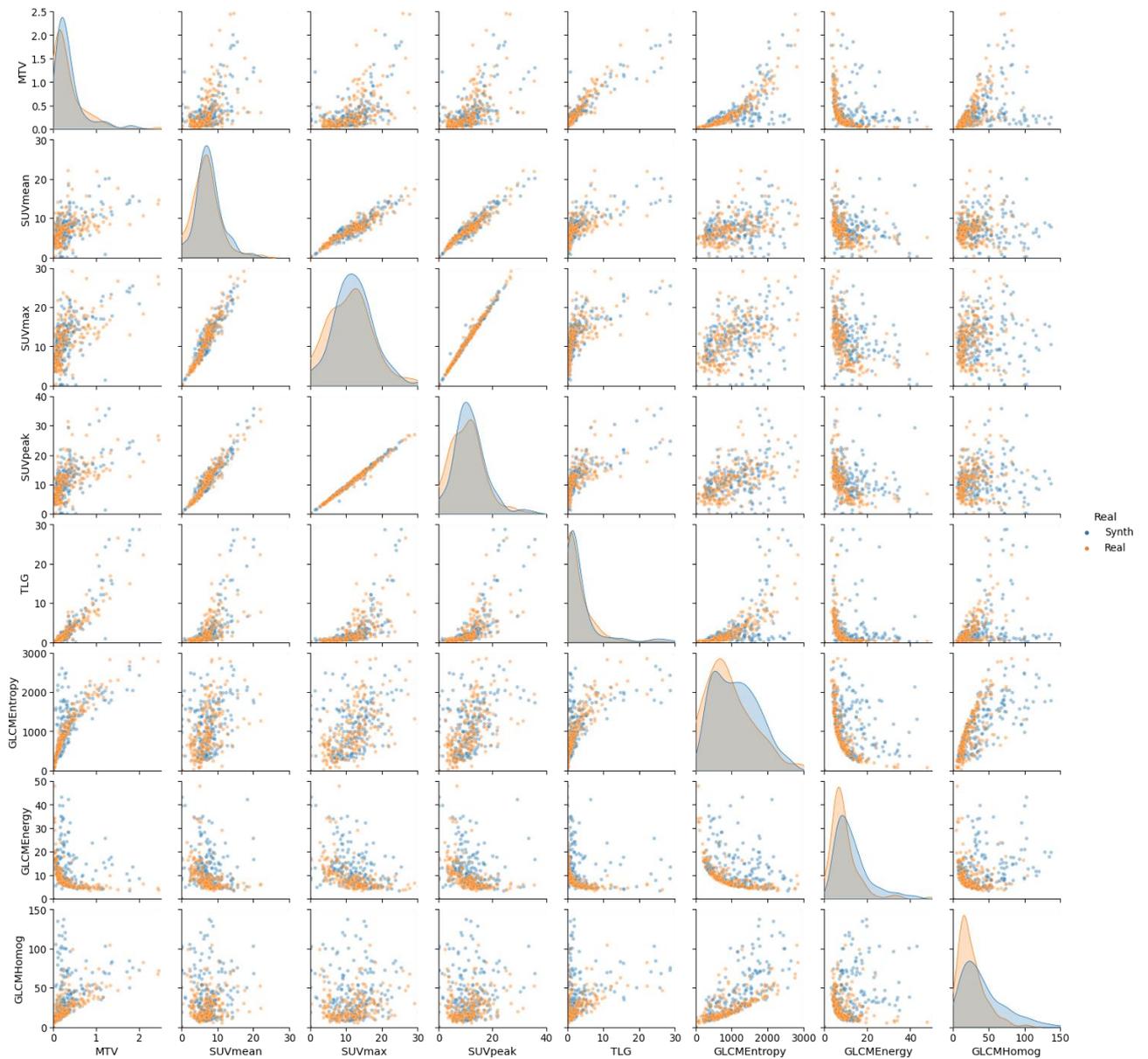

Figure 7. Pair plot of radiomic features calculated over segmented tumours for real (orange) and synthetic (blue) data. Synthetic data was generated by the **unconditional** TGAN.

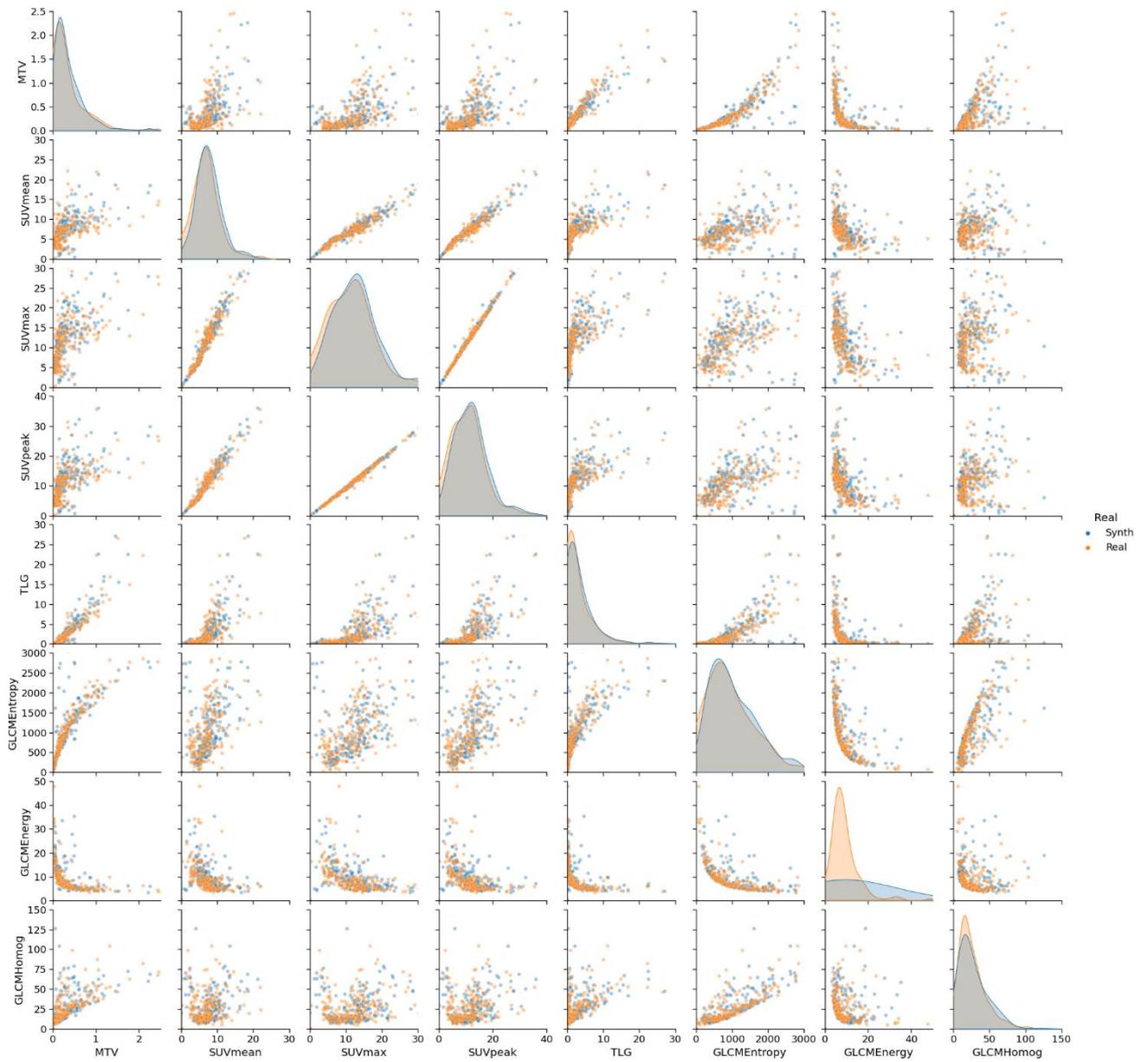

Figure 8. Pair plot of radionomic features calculated over segmented tumours for real (orange) and synthetic (blue) data. Synthetic data was generated by the **conditional** TGAN.

Table 1. Top row: Largest magnitude correlation coefficients for each feature calculated on real tumour segmentations. Middle row: Corresponding correlation coefficient calculated on the synthetic dataset. Bottom row: Percent differences.

| | **MTV** | **SUVmean** | **SUVmax** | **SUVpeak** | **TLG** | **GLCM Entropy** | **GLCM Energy** | **GLCM Homogeneity** |
|---|---|---|---|---|---|---|---|---|
| **Real Data** | GLCM Entropy r = 0.936 | SUVmax r = 0.966 | SUVpeak r = 0.997 | SUVmax r = 0.997 | MTV r = 0.936 | MTV r = 0.917 | GLCM Entropy r = -0.408 | GLCM Entropy r = 0.797 |
| **Synthetic Data** | r = 0.940 | r = 0.982 | r = 0.996 | r = 0.996 | r = 0.940 | r = 0.771 | r = 0.03 | r = 0.615 |
| **Δ%** | 0.4 | 1.7 | 0.1 | 0.1 | 0.4 | 15.9 | 93 | 23 |

## 4.4 Utility: Data Augmentation

Figure 9 shows the average Dice scores on the validation set during the training of the segmentation model on combinations of real and synthetic data. When validating the segmentation model on images from CHGJ (left figure), the synthetic data did not significantly improve segmentation performance. One explanation for this is may be that there are some properties inherent in images from CHGJ that are not present in other centers, and therefore additional synthetic data is not useful. When validating the segmentation model on images outside of CHGJ (right figure), we observed an increase in DSC when synthetic data was used.

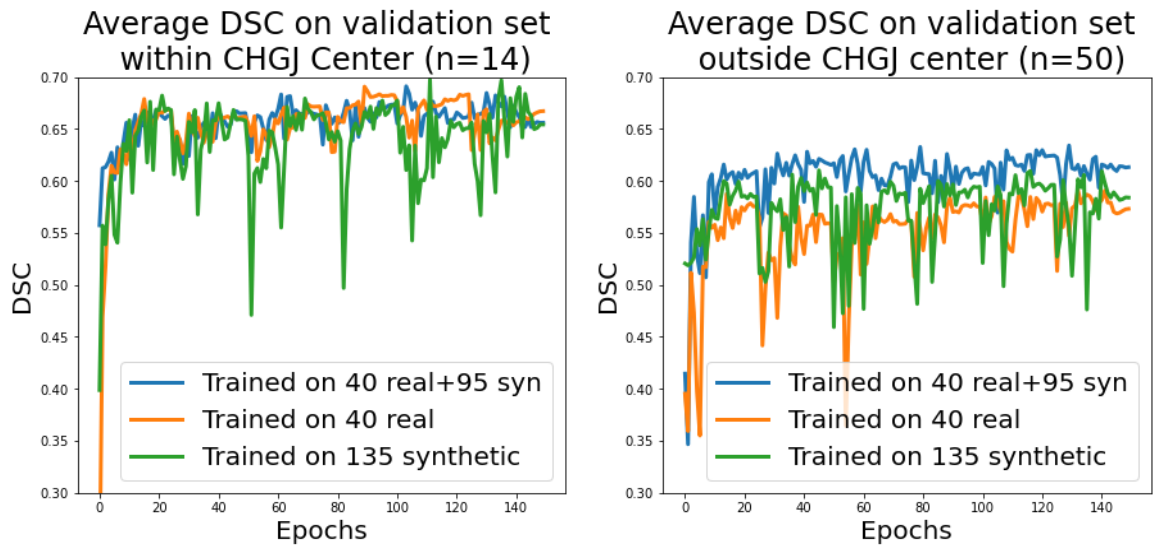

Figure 9. Utility Evaluation. Left: Adding synthetic data from other centers to the training set does not significantly affect DSC when validating on data from the same center. Right: Augmenting the model with synthetic data from other centers shows a clear improvement in segmentation accuracy when validating on data from outside CHGJ.

## 4.5 Discussion

3-D image generation using GANs is a very resource intensive task, given that a conventional 3-D GAN uses 3-D convolutional and deconvolutional layers that require a large amount of memory. This is partially addressed in the TGAN architecture by decomposing the generator into separate 2-D and 1-D generators. We successfully generated 3-D head and neck PET images with lesions using both unconditional and conditional TGANs. Our quantitative analysis shows that segmentation performance (based on the Dice score), as well as strong statistical correlations between radionomic features are preserved in the synthetic data set.

In this work, we chose to use segmentation as a measure of utility of the synthetic data set. We made this choice in part due to the public availability of high quality segmentation algorithms. However, there is also considerable research interest in PET disease classification. For example, distinguishing between Alzheimer's and Mild Cognitive Impairment (MCI)[19,20]. Our model could easily be adapted to generate Alzheimer's or MCI images to test utility in training classification algorithms as well.

One limitation of our work is that our quantitative analysis of real and synthetic datasets focused only on the properties of the segmented lesion. However, it is unclear what quantitative metrics should be used to measure TGAN's ability to generate the healthy anatomy outside the segmentation.

There are several planned improvements to our methodology in future works. For example, further improvements have been made to the TGAN architecture in a newer version, TGANv2[21]. One major improvement is the availability of multi-GPU capable code. Another major modification is made to the training and inference processes in TGANv2. The TGANv2 generator is able to output sparse samples for training at lower computational cost and dense samples for inference. Future work will test the capability of TGANv2 to generate higher resolution 3D medical images.

Other non-GAN based video generation techniques may be suitable for 3D image generation as well. For example, Yan et al. use vector quantized variational auto encoders to compress the video data and perform the generation in the compressed representation using transformers[22]. Further research is needed to assess how transformer based approaches compare to our approach for 3D image generation.

Our work seeks to facilitate sharing of medical data between researchers through generative models. It is well known that generative models can be vulnerable to membership inference attacks, in which an attacker can identify whether a data sample was used in the training of a model, which could in turn reveal an individual's private information[23]. Future work will assess our model's privacy protection in order to minimize this risk.

## 5. CONCLUSION

GAN-based models have shown promise in the field of medical image generation, especially for image-to-image translation tasks. Generating stand-alone 3D medical images, on the other hand, is a technically challenging and resource-intensive task that has not been well researched. We have shown that a video generation architecture, TGAN, can be adapted to generate 3D medical images that retain important image features and statistical properties of the training data set. We have shown that our model generalizes well and can produce realistic images with ground truth lesions based on user-input masks. This model may be beneficial to researchers who wish to train classification or segmentation algorithms, but lack a sufficient number of large labeled image datasets.

## REFERENCES

[1] van Panhuis, W.G., et al. "A systematic review of barriers to data sharing in public health". Bulletin of the World Health Organization, 88(6):468–468 (2010).
[2] Sheller, M.J., et al. Multi-institutional deep learning modeling without sharing patient data: A feasibility study on brain tumor segmentation. in International MICCAI Brainlesion Workshop. 2018. Springer.
[3] Zhao, Y., et al., Federated learning with non-iid data. arXiv preprint arXiv:1806.00582, 2018.
[4] Li, T., et al., Federated optimization in heterogeneous networks. arXiv preprint arXiv:1812.06127, 2018.
[5] Goodfellow, I., et al. [Generative Adversarial Nets]. Curran Associates, Inc., Advances in Neural Information Processing Systems, 27, 2672–2680 (2014).
[6] A. F. Frangi, S. A. Tsaftaris, and J. L. Prince. "Simulation and synthesis in medical imaging". IEEE Transactions on Medical Imaging, 37(3):673–679 (2018).
[7] Shin, H.-C., Tenenholtz, N.A., Rogers, J.K, Schwarz, C.G., Senjem, M.L, Gunter, J.L., Andriole, K., and Michalski, M. "Medical Image Synthesis for Data Augmentation and Anonymization using Generative Adversarila Networks." Lecture Notes in Computer Science 11037 (2018).

[8] Mahapatra, D., Bozorgtabar, B., Thiran, J.-P., and Reyes, M. "Efficient active learning for image classification and segmentation using a sample selection and conditional generative adversarial network," MICCAI, 580–588 (2018).
[9] Nie, D., Trullo, R., Petitjean, C., Ruan, S., Shen, D. "Medical image synthesis with context-aware generative adversarial networks," MICCAI 10435, 417–425 (2016).
[10] Islam, J., and Zang, Y. "GAN-based synthetic brain PET image generation." Brain Informatics, 7(1), 3 (2020).
[11] Yu, B., Wang, Y., Wang, L., Shen, D., and Zhou, L. [Medical Image Synthesis via Deep Learning]. Springer, Cham. Deep Learning in Medical Image Analysis. Advances in Experimental Medicine and Biology, vol. 1213 (2020).
[12] Wang, Y., Yu, B., Wang, L., Zu, C., Lalush, D.S., Lin, W., Wu, X., Zhou, J., Shen, D., and Zhou, L. "3D conditional generative adversarial networks for high-quality PET image estimation at low dose". NeuroImage 174:550–562 (2018).
[13] Bi, L., Kim, J., Kumar, A., Feng, D., and Fulham, M. [Synthesis of positron emission tomography (PET) images via multi-channel generative adversarial networks (GANs)]. Springer, Cham. Molecular imaging, reconstruction and analysis of moving body organs, and stroke imaging and treatment, 43–51 (2017).
[14] Vondrick, C., Pirsiavash, H., and Torralba, A. "Generating Videos with Scene Dynamics". NIPS, (2016).
[15] Saito, M., Matsumoto, E., and Saito, S. "Temporal Generative Adversarial Nets with Singular Value Clipping". ICCV, (2017).
[16] Iantsen A., Visvikis, D., and Hatt, M. "Squeeze-and-Excitation Normalization for Automated Delineation of Head and Neck Primary Tumors in Combined PET and CT Images". Springer, Cham. Head and Neck Tumor Segmentation. HECKTOR 2020. Lecture Notes in Computer Science, 12603 (2021).
[17] Cook, G., Azad, G., Owczarczyk, K., Siddique, M., and Goh, V. Challenges and Promises of PET Radiomics. International journal of radiation oncology, biology, physics 102(4):1083–1089 (2018).
[18] Andrearczyk, V., Oreiller, V. and Depeursinge, A. eds., 2021. Head and Neck Tumor Segmentation: First Challenge, HECKTOR 2020, Held in Conjunction with MICCAI 2020, Lima, Peru, October 4, 2020, Proceedings (Vol. 12603). Springer Nature.
[19] Yan, W., Zhang, Y., Abbeel, P., and Srinivas, A. "VideoGPT: Video Generation using VQ-VAE and Transformers." Arxiv, 20 April 2021, <https://arxiv.org/pdf/2104.10157.pdf> (07 July 2021).
[20] Cabral, C., and Silveira, M. "Alzheimer's Disease Neuroimaging Initiative. Classification of Alzheimer's disease from FDG-PET images using favourite class ensembles". Annu Int Conf IEEE Eng Med Biol Soc. 2477-2480 (2013).
[21] Saito, M., Saito, S., Koyama, M., and Kobayashi, S. "Train Sparsely, Generate Densely: Memory-efficient Unsupervised Training of High-resolution Temporal GAN". International Journal of Computer Vision. 128:2586-2606 (2020).
[22] Singh, S., Srivastava, A., Mi, L., Caselli, R. J., Chen, K., Goradia, D., Reiman, E. M., and Wang, Y. Deep Learning based Classification of FDG-PET Data for Alzheimers Disease Categories. Proceedings of SPIE--the International Society for Optical Engineering, 10572 (2017).
[23] Shokri, R., Stronati, M., Song, C,. and Shmatikov, V. Membership Inference Attacks Against Machine Learning Models. IEEE Symposium on Security and Privacy (S&P), Oakland. (2017). 10.1109/SP.2017.41.